\documentclass[longauth,letter]{aa}  

\usepackage[T1]{fontenc} 
\usepackage{siunitx}
\usepackage{amsmath}
\usepackage{amssymb}
\usepackage{graphicx}
\usepackage{longtable}   
\usepackage{lscape}
\usepackage[export]{adjustbox}
\usepackage{natbib}
\usepackage{subcaption}
\usepackage{calrsfs}
\usepackage{array}
\usepackage{hyperref}

\bibpunct{(}{)}{;}{a}{}{,}

\newcommand{\safir}{S{\small AFIR}}

\newcommand{\harps}{H{\small ARPS}}
\newcommand{\hipp}{Hipparcos}

\newcommand{\sphere}{S{\small PHERE}}
\newcommand{\naco}{N{\small aCo}}

\begin{document} 

   \title{Constraining the properties of HD 206893 B}

   \subtitle{A combination of radial velocity\thanks{\harps \ run 089.C-0739(A), 192.C-0224(C), 099.C-0205(A), 098.C-0739(A) and 1101.C-0557(A)}, direct imaging\thanks{\sphere \ run 096.C-0388, 097.C-0865(D) and 099.C-0708(A).} , and astrometry\thanks{\cite{DR2A1}} data}

   \author{A. Grandjean
          \inst{1}
          \and
          A.-M. Lagrange\inst{1}
        \and
          H. Beust\inst{1}
        \and
          L. Rodet\inst{1}
        \and
      J. Milli\inst{2}
        \and
        P. Rubini\inst{3}
        \and
        C. Babusiaux \inst{1,} \inst{4}
        \and
        N. Meunier\inst{1}
        \and
        P. Delorme \inst{1}
        \and
        S. Aigrain \inst{5}
        \and
        N. Zicher  \inst{5}
        \and
        M. Bonnefoy \inst{1}
        \and    
        B. A. Biller \inst{6,} \inst{15}
        \and
        J.-L. Baudino \inst{7}
        \and
        M. Bonavita \inst{6}
        \and
        A. Boccaletti \inst{8}
        \and
        A. Cheetham \inst{9}
        \and 
        J. H. Girard \inst{10}
        \and 
        J. Hagelberg \inst{9}
        \and
        M. Janson \inst{11}
        \and
        J. Lannier \inst{1}
        \and
        C.  Lazzoni \inst{12}
        \and
        R. Ligi \inst{13}
        \and
        A.-L. Maire \inst{14}
        \and
        D. Mesa \inst{15}
        \and
        C. Perrot \inst{8,} \inst{16,} \inst{17}
        \and
        D. Rouan \inst{8}
        \and
        A. Zurlo \inst{18,} \inst{19,} \inst{20}
          }

   \institute{
Univ. Grenoble Alpes, CNRS, IPAG, 38000 Grenoble, France
 \\
\email{Antoine.Grandjean1@univ-grenoble-alpes.fr}
\and
European Southern Observatory (ESO), Alonso de C\'ordova 3107, Vitacura, Casilla 19001, Santiago, Chile
\and
Pixyl S.A. La Tronche, France
\and
GEPI, Observatoire de Paris, Universit\'{e} PSL, CNRS, 5 Place Jules Janssen, 92190 Meudon, France
\and
Department of Astrophysics, University of Oxford
\and 
Institute for Astronomy, University of Edinburgh, Blackford Hill, Edinburgh EH9 3HJ, UK ; Centre for Exoplanet Science, University of Edinburgh, Edinburgh, UK
\and 
Department of Physics, University of Oxford
\and
LESIA, Observatoire de Paris, PSL Research University, CNRS, Sorbonne Universités, UPMC Univ. Paris 06, Univ. Paris Diderot
\and 
Observatoire de Genève, University of Geneva, 51 Chemin des Maillettes, 1290, Versoix, Switzerland
\and
Space Telescope Science Institute, Baltimore, United States
\and
Department of Astronomy, Stockholm University, AlbaNova University Center, 106 91 Stockholm, Sweden
\and
INAF - Osservatorio Astronomico di Padova
\and
INAF–Osservatorio Astronomico di Brera, Via E. Bianchi 46, I-23807, Merate, Italy
\and
Max-Planck-Institut für Astronomie, Königstuhl 17, 69117 Heidelberg, Germany
\and
Astronomical Observatory of Padova/Asiago
\and
 Instituto de F\'isica y Astronom\'ia, Facultad de Ciencias, Universidad de Valpara\'iso, Av. Gran Breta\~na 1111, Valpara\'iso, Chile
\and
 N\'ucleo Milenio Formaci\'on Planetaria - NPF, Universidad de Valpara\'iso, Av. Gran Breta\~na 1111, Valpara\'iso, Chile
\and
Aix Marseille Univ, CNRS, CNES, LAM, Marseille, France
\and
Núcleo de Astronomía, Facultad de Ingeniería y Ciencias, Universidad Diego Portales, Av. Ejercito 441, Santiago, Chile
\and
Escuela de Ingeniería Industrial, Facultad de Ingeniería y Ciencias, Universidad Diego Portales, Av. Ejercito 441, Santiago, Chile
           }

   \date{Received 2019 january 11  / Accepted 2019 mai 24}

 
  \abstract
   {High contrast  imaging enables the determination of orbital parameters for substellar companions (planets, brown dwarfs) from the observed relative astrometry and the estimation of model and age-dependent masses from their observed magnitudes or spectra. Combining  astrometric positions with radial velocity gives direct constraints on the orbit and on the dynamical masses of companions.
A  brown dwarf was discovered with the VLT/\sphere \ instrument at the Very Large Telescope (VLT) in 2017, which orbits at $\sim \SI{11}{au}$ around HD 206893. Its mass was estimated between $12$ and $\SI{50}{M_J}$ from  evolutionary models and its photometry. However, given the significant uncertainty on the  age of the system and the peculiar spectrophotometric properties of the companion, this mass is not well constrained.

}
   {We aim at constraining the orbit and dynamical mass of HD 206893 B.}
   {We combined radial velocity data obtained with \harps \ spectra and astrometric data obtained with the high contrast imaging VLT/\sphere \  and VLT/\naco \ instruments, with a time baseline less than three years. We then combined those data with astrometry data obtained by \hipp \ and Gaia with a time baseline of $24$ years.
We used a Markov chain Monte Carlo  approach to estimate the orbital parameters and dynamical mass of the brown dwarf from those data.}
   {We infer a period between $21$ and $\SI{33}{yr}$ and an inclination in the range $20-\SI{41}{\degree}$ from pole-on from HD 206893 B 
relative astrometry.
The RV data show a significant RV drift over $\SI{1.6}{yr}$.
We show that HD 206893 B cannot be the source of this observed RV drift as it would lead to a dynamical mass inconsistent with its photometry and spectra and with \hipp \ and Gaia data.
An additional inner (semimajor axis in the range $1.4-2.6 \ \si{au}$) and massive ($\sim \SI{15}{M_J}$) companion is needed to explain the RV drift, which is compatible with the available astrometric data of the star, as well as with the VLT/\sphere \ and VLT/\naco \ nondetection.

}
   {}


   \maketitle

%

\section{Introduction}

Owing to their wide separations from their host stars ($> \SI{20}{au}$), dynamical masses of directly imaged planet and brown dwarf (BD) companions are difficult to measure.  Thus, most of the current mass estimates for these companions rely on evolutionary models.
These models still require qualibration, especially at young ages.
Such calibrations can be performed using systems for which the companion mass is independently measured. 

Indirect measurements of the host star can also provide important constraints on companion properties.  In particular, measuring the radial velocity (RV) of a host star enables measurement or at least a constraint on the dynamical mass, $M_c \sin{i}$, of a companion.
\cite{Janson}, \cite{Snellen}, and  \cite{Trent}  demonstrated that \hipp \ and Gaia \ measurements of the proper motion of a host star can also be used to better constrain the dynamical mass of its companion .

HD 206893 is a nearby F5V  star, located at $40.81^{+0.11}_{-0.11} \si{pc}$ \citep{DR2A1}. 
\cite{Zuckerman} estimated an age between  $0.2$ and $\SI{2}{\giga yr}$, which was later refined to $250^{+450}_{-200} \ \si{\mega yr}$ by \cite{Delorme}  using \cite{Desidera} method.
Additional characteristics of the stars are presented in Table \ref{tab_206}.

\cite{Milli} discovered a BD companion around HD 206893 via direct imaging using the High-contrast Exoplanet REsearch (SPHERE) instrument  \citep{SPHERE_Beuzit} and the \naco \ instruments at the Very Large Telescope (VLT).  
The companion has a projected separation of $\SI{270}{\milli as}$ (\SI{11}{au}). The star also appears to host a barely resolved disk inclined by  $40 \pm 10 \si{\degree}$ from pole-on.
From their data taken in 2015-2016, \cite{Milli} estimated a period of $\sim \SI{37}{ yr}$, in the case of low eccentricity, and the mass of the companion from models in the range $24$ to $\SI{73}{M_J}$, for a corresponding age in the range  $0.2$ to $\SI{2}{\giga yr}$. 
Using these data and VLT/\sphere \ epochs taken in 2016, \cite{Delorme} refined the orbital parameters, finding a period $P$ of  $\sim \SI{27}{yr}$ and mass $M_c$ between $12$ and $\SI{50}{M_J}$ for a corresponding age in the range  $50-\SI{700}{\mega yr}$.
As the age of the system is poorly known, the estimated mass of HD 206893 B, derived from brightness-mass relations, is still poorly constrained. 

We report on the High Accuracy Radial velocity Planet Searcher  (\harps) observations of HD 206893, which reveal an RV drift and two new VLT/\sphere \ datasets.
The data are presented in Section \ref{Data}, and the analysis of the RV is presented in Section \ref{RV_analyse}.
In Section \ref{POS} we use high contrast direct imaging  observations to constrain the orbit of HD 206893 B.
In Section \ref{DI_RV} we combine high contrast direct imaging and RV observations to constrain the orbit and dynamical mass of HD 206893 B.
In Section \ref{RV_ID_HG} we combine high contrast direct imaging, RVs, and proper motion data of the host star from \hipp \ and Gaia (presented in Section \ref{g_h}) to fully constrain the system.
Finally, we discuss our results in Section \ref{discussion}.

\begin{table}
\center
\begin{tabular}[h!]{lcl}
\hline
Parameter & Value & Ref \\ \hline
$V \  (mag) $ & $6.67$  & Tycho-2$^1$\\
$B - V \  (mag)$ & $0.44$ & Tycho-2$^1$ \\
$Parallax \ (\si{\milli as})$ & $24.51 \pm 0.06$ & \emph{Gaia} DR2$^2$ \\
$\mu_{\alpha} \ (\si{\milli as \per yr}$) & $93.78 \pm 0.09$ & \emph{Gaia} DR2$^2$ \\
$\mu_{\delta} \ (\si{\milli as \per yr}$) & $0.016 \pm 0.08$ &  \emph{Gaia} DR2$^2$ \\
$M_* \ (M_{\odot})$ & $1.32 \pm 0.02$ & Delorme$^3$\\
$i_* \ (\si{\degree})$ & $ 30 \pm 5$ &  Delorme$^3$ \\
$T_{eff} \ (\si{\kelvin})$ & $ 6500 \pm 100$ &  Delorme$^3$ \\
$Age \ (\si{\mega yr})$ & $250_{-200}^{+450}$ &  Delorme$^3$\\
$RV  \ (\si{\kilo \meter \per \second})$ & $-12.45$ & \emph{Gaia} DR2$^2$  \\
$V \sin i \ (\si{\kilo \meter \per \second})$ & $29$ & This work \\
$logR_{HK} $ & $-4.65 \pm 0.02$ & This work \\

\end{tabular}
\caption{HD 206893 stellar parameters.  1 \cite{Tycho-2}, 2 \cite{DR2A1}, 3  \cite{Delorme}}
\label{tab_206}
\end{table}

\begin{figure*}[t!]
\captionsetup[subfigure]{labelformat=empty}
  \centering
\begin{subfigure}[b]{0.45\textwidth}
\includegraphics[scale=0.37]{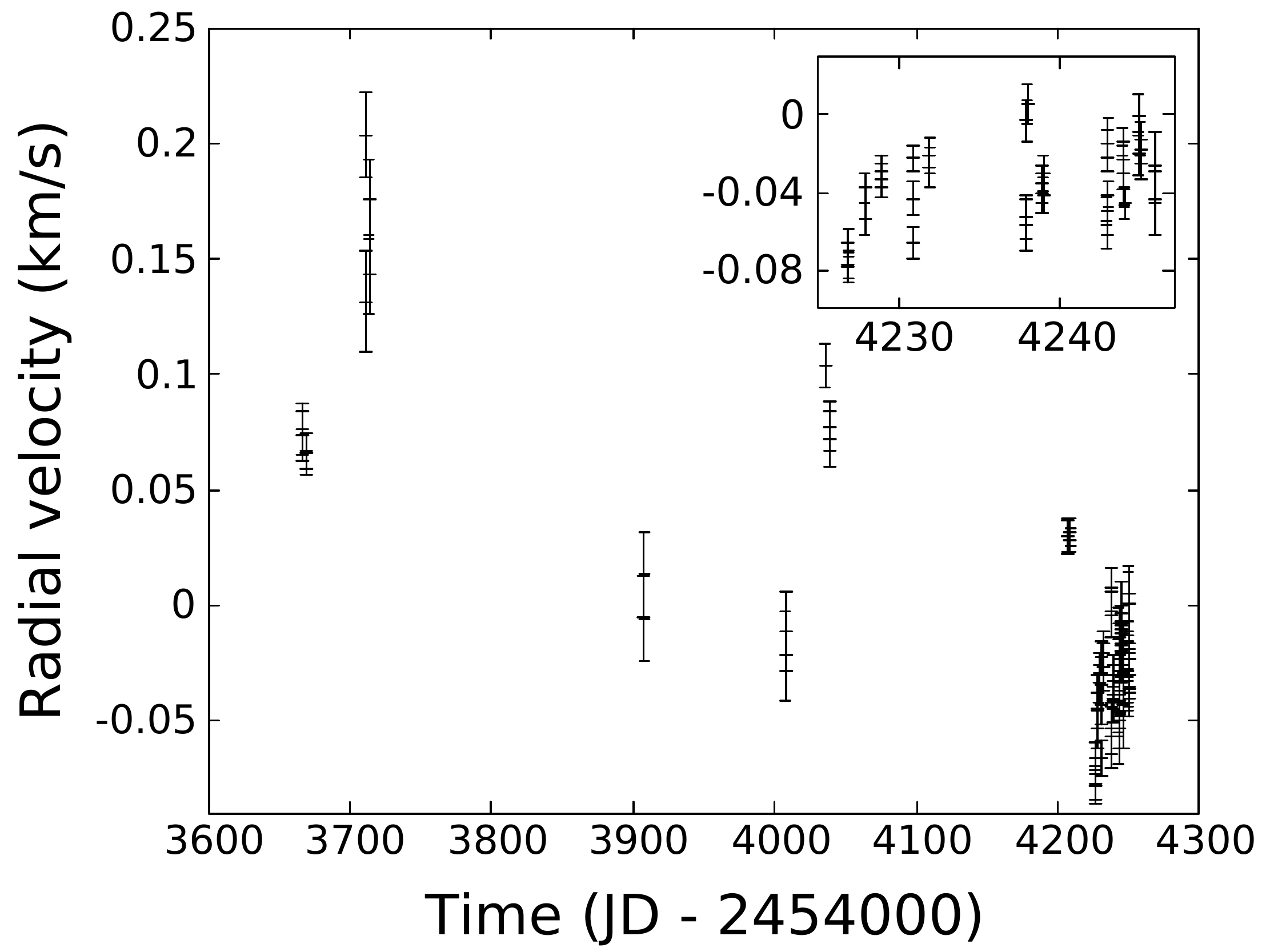}
\caption{ \label{fig_RV}}
\end{subfigure}
\begin{subfigure}[b]{0.45\textwidth}
\includegraphics[scale=0.37]{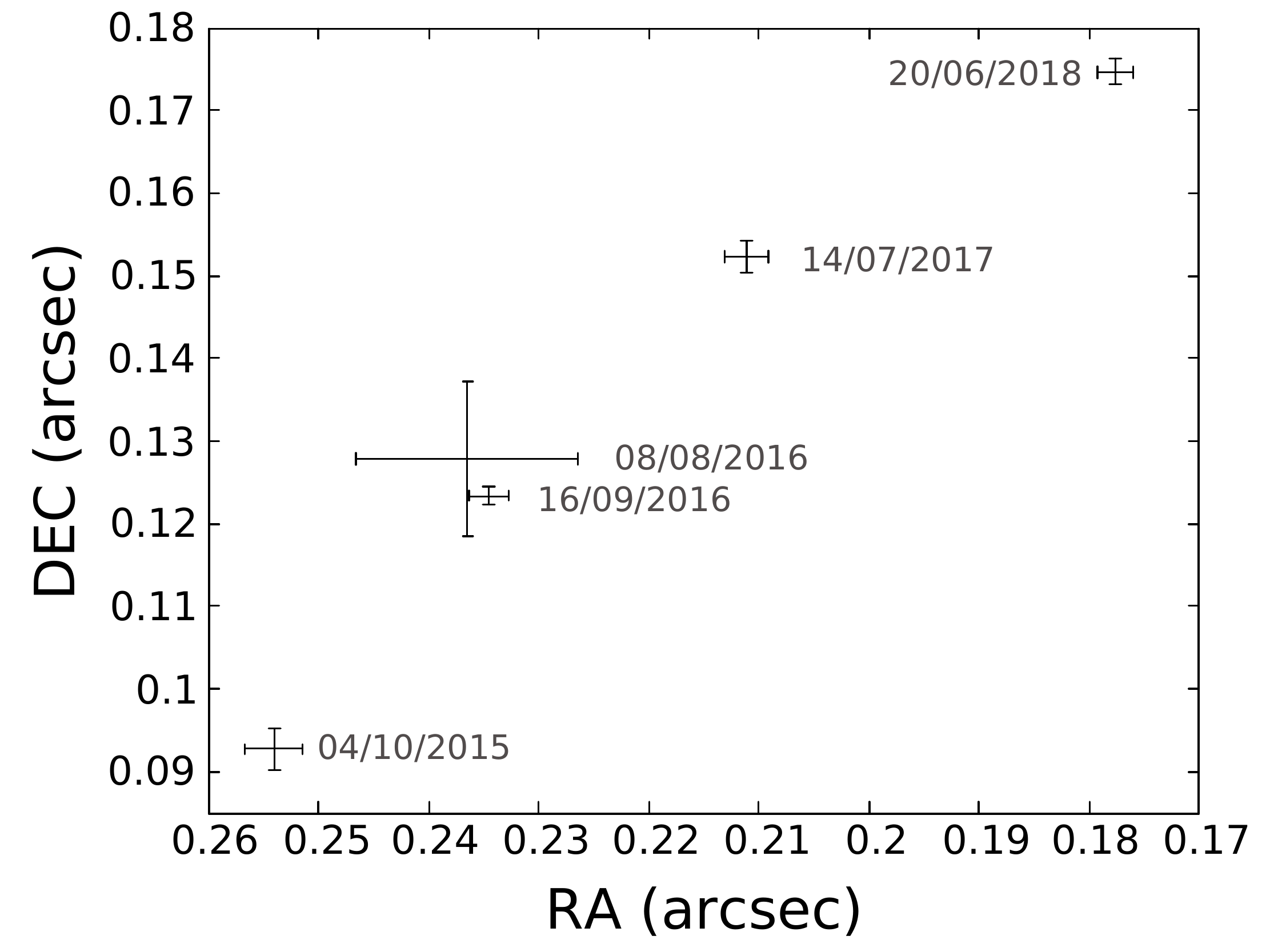}
\caption{ \label{fig_ID}}
\end{subfigure}
\caption{\emph{Left}: \harps \ RV variations of HD 206893 over a $700$ day period, \emph{Top right corner}: Zoom on short-term variations. \emph{Right}: Position of HD 206893 B relative to the host star, as measured with VLT/\sphere \ and VLT/\naco \ at different epochs  \label{RV_ID}} 
\end{figure*}

\section{Data}

\label{Data}

        \subsection{Radial velocities}
\label{RV}

HD 206893 was observed over a 1.6 year timespan, between October 2016 and May 2018 with \harps \ (La Silla, ESO) \citep{HARPS}, as part of our Young Nearby Stars survey (YNS; \cite{Lagrange}). 
A total of 66 spectra were obtained with a sun altitude below $-10$ degrees. We use our Spectroscopic data via analysis of the Fourier Interspectrum RVs. (\safir) software \citep{SAFIR}  to compute the RVs, which are shown in Figure \ref{fig_RV}.

        \subsection{Direct imaging astrometry}

\label{DI}

In this paper, we used all epochs obtained for HD 206893 with VLT/\sphere \ in Infra-Red Dual Imaging and Spectrograph (IRDIS) mode and with VLT/\naco.
\cite{Milli} obtained the first epochs in 2015 and 2016. Then, as part of the The SpHere INfrared survey for Exoplanets (SHINE) survey \citep{chauvin}, HD 206893 was observed with \sphere \ in September 2016, whose results are reported in \cite{Delorme}.
Finally, HD 206893 was observed two more times with \sphere \ in 2017, as part of the follow-up of the \sphere \ High-Angular Resolution Debris Disk Survey (SHARDDS; \cite{Milli}), and in 2018, as part of the SHINE survey.

 The data were reduced and calibrated as described in \cite{Maire_2016} and \cite{Delorme}. Table \ref{tab_ID} summarizes the observational set up and the measured  positions of the companion. The companion positions are presented in Figure \ref{fig_ID}. The last epoch shows a curvature of the trajectory, which is promising for further orbital characterization of the companion.

\begin{table}
\center
\resizebox{0.5\textwidth}{!}{ 
\begin{tabular}[h!]{ll|cc}
\hline
Date & Instrument &Separation  & Position angle  \\ 
& and filter & $(\si{\milli as})$ & $(\si{\degree})$ \\\hline
2015-Oct-04 &\sphere \   H& $270.4 \pm 2.6$ & $69.95 \pm 0.55$ \\
2016-Aug-08 & \naco \ L'& $268.8 \pm 10.4$ & $61.6 \pm 1.9$ \\
2016-Sep-16 &\sphere \  K1/K2 & $265 \pm 2$ & $62.25 \pm 0.1$ \\
2017-Jul-14& \sphere \  H & $260.3 \pm 2$ & $54.2 \pm 0.4$ \\
2018-Jun-20 &\sphere \  H2/H3 & $249.11 \pm 1.6$ & $45.5 \pm 0.37$  \\
\end{tabular}}
\caption{Astrometry: separation and position angle of HD 206893 B.}
\label{tab_ID}
\end{table}

\subsection{\hipp \ and Gaia proper motion}

\label{g_h}

Proper motion and position measurements for HD 206893 are available from both the \hipp \ \citep{hipp2} and Gaia \citep{DR2A1} missions, with a time baseline of 24 years between the two missions.  

\hipp \ and Gaia probe the position of the photocenter of the system, which is not the same as the barycenter of the system.
Once corrected from the difference in position of the star induced by the companion, the proper motion deduced from the difference of position between  \hipp \  and Gaia measurements corresponds to the proper motion of the barycenter of the system.
The difference between the proper motion as measured by \hipp \ and Gaia and the barycenter proper motion provides information on the variation of the tangential velocity over the duration of both datasets (two years for Hipparcos, more than two years for Gaia).

These measurements thus constrain both the orientation of the orbit and the dynamical mass of the companion because the impact of the companion on the proper motion of the star should be within the uncertainties of the measurements.
\cite{Snellen} and \cite{Trent} have provided an example of such constraints gained by combining \hipp \ and Gaia measurements with direct imaging and RV constraints for the planet $\beta$ Pic b.

\begin{table}[h!]
\center
\begin{tabular}[h!]{c|cccc}
& $\mu^{RA}$ &$\sigma_{\mu}^{RA}$ & $\mu^{DEC}$& $\sigma_{\mu}^{DEC}$ \\
&\multicolumn{2}{c}{($\si{\milli as \per yr}$)}& \multicolumn{2}{c}{($\si{\milli as \per yr}$)} \\ \hline \hline
\hipp \ & $93.6$ & $0.66$ & $0.4$ & $0.37$   \\ 
\hipp-Gaia  & $94.21$ & $0.02$ & $0.17$ & $0.014$\\  
Gaia & $93.8$ & $0.1$ & $-0.11$ & $0.09$ \\ 
\end{tabular}
\caption{\hipp-2 \  and Gaia DR2 proper motion  and proper motion deduced from the difference of position.}
\label{Table_pm}
\end{table}

Both the Hipparcos and Gaia five-parameter solutions are well behaved for HD 206893 (see Appendix \ref{corr_gaia}). 
We present the proper motion measured by \hipp \ and Gaia and the proper motion deduced from the difference of position in  Table \ref{Table_pm}.

\section{Radial velocity analysis}
\label{RV_analyse}

The RVs show a strong dispersion due to high frequency variations (hereafter HF). In addition, a linear trend is clearly present (see Figure \ref{RV_ID}).

\begin{figure}[h]
\centering
\includegraphics[scale=0.3]{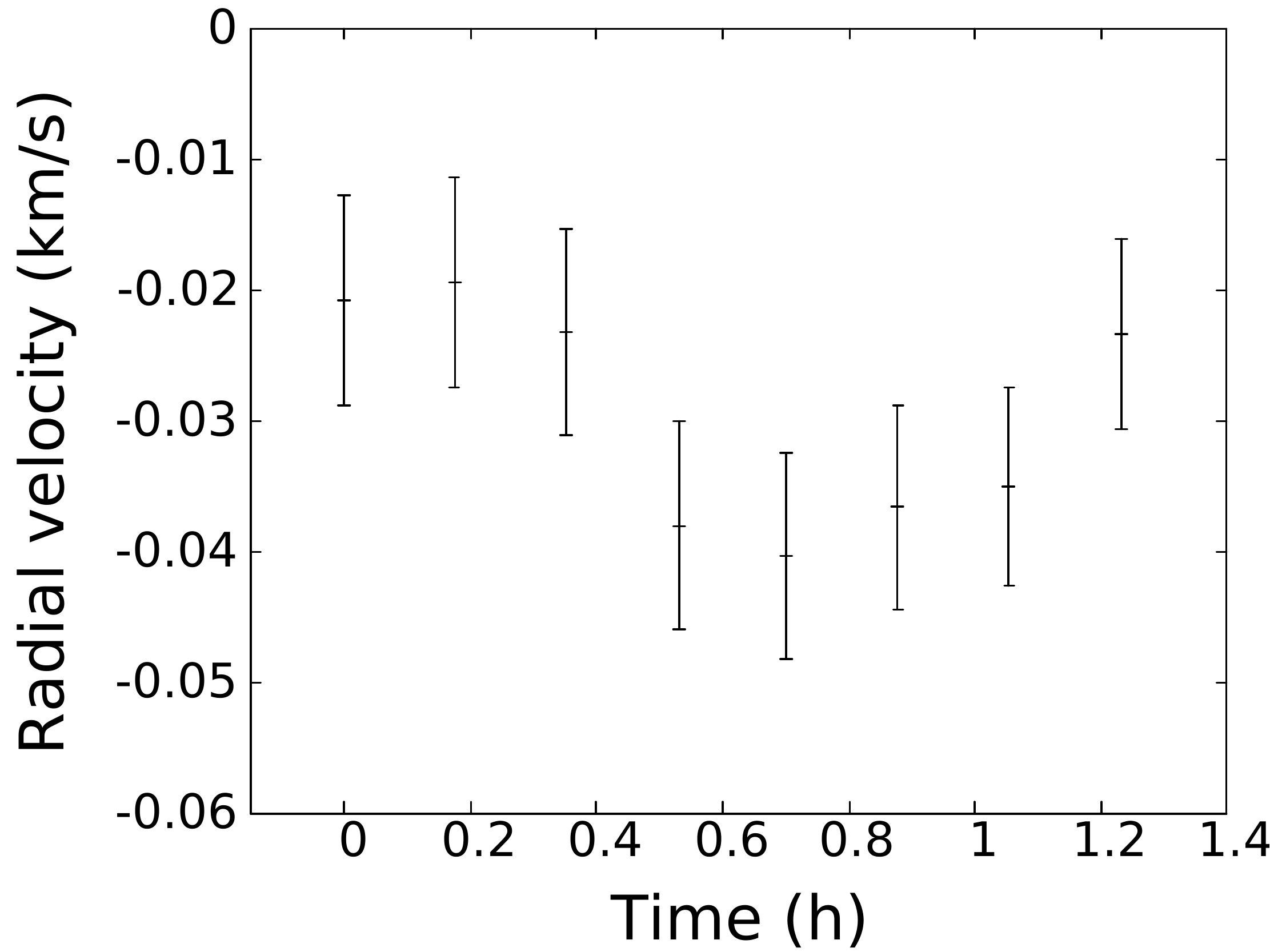}
\caption{HD 206893 RV variations on JD 2458249. \label{last_rv}}
\end{figure}

The HF variations seem to have periods slightly less than one day (see Figure \ref{fig_RV}), which is consistent with the photometric variations reported in \cite{Delorme}.
Given the star's spectral type of F5V, the variations could a priori either be due to pulsations or spots. 
The bisector velocity span (BVS) \citep{2001Queloz_bis} versus RV diagram shows a vertical spread on the 2018 observations and our last epoch shows sinusoidal variations over one hour (see Figure \ref{last_rv}). Such a short timescale  indicates that the HF variations are due to pulsations rather than magnetic activity. Additionally, vertical (BVS,RV) spread also indicates pulsations rather than spots because spots usually produce a correlation between RV and BVS (see examples in \cite{Lagrange_2009, Lagrange}).

A linear regression gives a slope of $\SI{92}{\meter\per\second\per yr}$ for the trend. However, because of the very poor data sampling in 2017, coupled with high amplitude stellar variations, the slope is not well constrained. Binning the RV would also lead to a poor estimation of the slope.

\begin{figure}[t!]
\centering
\includegraphics[scale=0.4]{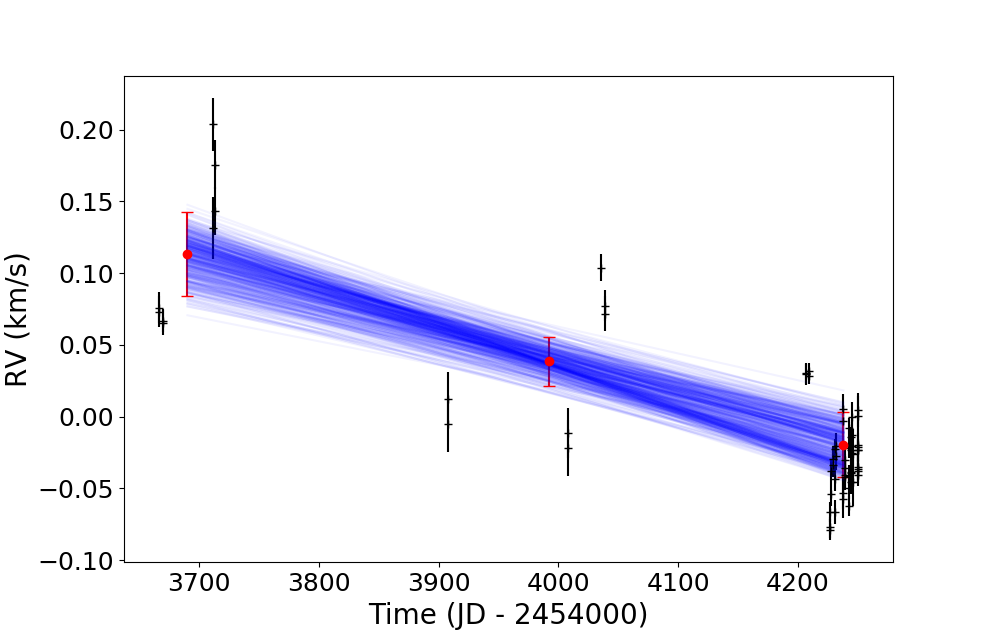}
\caption{RV drifts solutions deduced from the fit of the pulsations (\emph{blue}) and  deduced RV  (\emph{red}).  The original RV are plotted in black.\label{puls_fit}}
\end{figure}

To better constrain the slope of the trend, we used a genetic algorithm that fits the RV variations with a sum of 3 sine functions with periods between $0.5$ an $2$ days to fit the pulsations plus a linear slope  to fit the long-term RV variations and an offset to account for the fact that the RV are relative to a reference spectrum \citep{SAFIR}.
Fitting three sine function components provides a good overall fit without over-fitting.
We used 500 different seeds to estimate the uncertainties associated with the fit. We obtained a median slope of $87^{+16}_{-14} \ \si{\meter\per\second\per yr}$ and a median offset of $1^{+0.17}_{-0.15} \ \si{\kilo\meter\per\second}$. 
Given the resulting $(offset,slope)$ distribution, we can associate to each date a distribution of RV with a linear model. We present this RV distribution in Figure \ref{puls_fit}.
Our RV observations fit broadly into three distinct time epochs. For each of these epochs, we estimated the corresponding RV of the drift and its $1-2 \sigma$ errorbar from the median of the RV distribution and its $1-2 \sigma$ limits at the mean date of the  epoch.
 The resulting RVs are presented in red in Figure  \ref{puls_fit}. We use these RVs in the rest of the study.

\label{fit_puls}

\section{Direct imaging astrometry}

\label{POS}

To constrain the orbital parameters of the BD we fit the astrometry of the companion using a Bayesian approach. We used the model described in \cite{Model_POS_RV} based on  the six orbital parameters to compute the astrometric position of the companion. This model was implemented into the likelihood function used by the python implementation of the Affine Invariant Markov chain Monte Carlo (MCMC) Ensemble sampler \emph{emcee} \citep{emcee,emcee_2010}.
We describe the parameters and our priors in Appendix \ref{desc_pos}.

We present the results of the MCMC run in Figure \ref{corner_pos}, and the $1-2 \sigma$ intervals in Table \ref{tab_pos}. 
At the end of the MCMC run, the median reduced $\chi^2$ of the chain is $1.1$, with a standard deviation of $0.4$, ensuring that the data are well fit by the model.
We find a period consistent within $1 \sigma$ with $\SI{27}{yr}$ period estimated by \cite{Delorme}; in our results $P$ is in the range $21-\SI{33}{yr}$.
The orbit is most likely eccentric, between $0$ and $0.5$, and has a maximum of probability at $0.37$.
We find an inclination in the range $20-\SI{41}{\degree}$ from pole-on, which is consistent with the inclination of the disk, i.e., $i_{Disk} = 40 \pm 10 \si{\degree}$, and the inclination of the equatorial plane of the star, i.e., $i_* = 30 \pm 5 \si{\degree}$. The orbit of the BD is then likely coplanar with the disk and in the equatorial plane of its host star.
We note a  strong degeneracy between period and inclination.  We also find a degeneracy between inclination and eccentricity and between the period and eccentricity.
We do not discriminate between extended and inclined circular orbits and shorter, eccentric and close to pole-on orbits.

\begin{table}[h!]
\center
\begin{tabular}[h!]{c|c c | cc |}
Parameter & \multicolumn{2}{c|}{$1 \sigma $ interval}  &  \multicolumn{2}{c|}{$2 \sigma $ interval}\\ \hline
$P \ (\si{yr})$ & $21$ & $33$ & $19$ & $41$\\
$TP \ (\si{yr})$ & $-7$ & $-0.3$ & $-14$ & $7.7$\\
$e $ & $0.08$ & $0.38$ & $0.01$ & $0.42$\\
$i \ (\si{\degree})$ & $139$ & $160$ & $134$ & $171$\\
$\Omega \ (\si{\degree})$ & \multicolumn{4}{c|}{Max at $255 \ mod  \ 180$}  \\
$\omega \ (\si{\degree})$ &  \multicolumn{4}{c|}{Max at $55 \ mod \ 180$}  \\
\end{tabular}
\caption{$1 \sigma$ and $2 \sigma$ intervals of the orbital parameters of HD 206893 B estimated by the MCMC  on the companion astrometry alone.}
\label{tab_pos}
\end{table}

\section{Direct imaging astrometry and RV analysis}

\label{DI_RV}
To constrain the dynamical mass of the BD, we assumed that the observed companion was responsible for the RV drift then fit the combined RV, presented in Section \ref{RV_analyse}, and direct imaging data with a MCMC. Our approach is similar to that used by \cite{Model_POS_RV} on $\beta$ Pic b.  We used the six orbital parameters, the mass of the BD, and an offset in the RV in our orbital fit. This model was implemented into the likelihood function of our MCMC.
We describe the parameters and our priors in Appendix \ref{desc_pos_rv}.
\label{corr}

We present the results of the MCMC run in Appendix \ref{result_pos_rv}. 
At the end of the MCMC run, the median reduced $\chi^2$ of the chain is $1.1$, with a standard deviation of $0.4$, ensuring that the data are well fit by the model.
We find similar orbital parameters compared to the fit on the astrometric position of the companion alone. The dynamical mass is now constrained, essentially between $60$ and $\SI{720}{M_J}$ ($2 \sigma$) with a maximum of probability at $\SI{140}{M_J}$. This mass is incompatible at $2 \sigma$ with the $12-\SI{50}{M_J}$ mass inferred from its observed photometry and spectra by \cite{Delorme}.

\section{Direct imaging astrometry, RV, and star proper motion analysis}

The proper motion induced by HD 206893 B during \hipp \ and Gaia observations should be consistent with the difference between the proper motion measured by \hipp \ and Gaia and the barycenter proper motion.  
The \hipp \ and Gaia measurements provide mass constraints complementary to those from the RV observations, as  the observed proper motion traces the tangential velocity of the star induced by the BD.
\hipp \ and Gaia astrometry also yields information on the orientation of the orbit and is thus complementary to direct imaging astrometry as well.

None of the solutions at the end of the MCMC on the companion relative astrometry and RV fit \hipp \ or Gaia measurements in right ascension and declination, as the amplitude of the tangential velocity is too high (see Figure \ref{tg_pos_rv}).

We add the model to compute the proper motion induced by the companion and the reference proper motion in the likelihood function of the MCMC (see Appendix \ref{hg_model} and \ref{ref_pm}). We describe the parameters and our priors in Appendix \ref{desc_pos_rv_hg}. Then, we run the MCMC with the direct imaging astrometric data, the three RV epochs presented in Section \ref{RV_analyse} and the proper motion measured by \hipp \ and Gaia.
The results are presented in Appendix \ref{result_pos_rv_h_g_hg}.
At the end of the MCMC run, the median reduced $\chi^2$ of the chain is $3.4$; the standard deviation is $0.5,$ which is significantly worse than the results on relative astrometry and RV. The median $\chi^2$ on the RV data is six times higher than for the solutions of the MCMC on the direct imaging astrometry and RV.
We find a mass of $10^{+5}_{-4} \  \si{M_J}$, which is consistent at $1\sigma$ with the mass inferred from the photometry and spectra of HD 206893 B by \cite{Delorme}; this result favors the hypothesis of young ages.
This indicates that  \hipp \  and Gaia are probing the  dynamical influence of the BDs but not the RVs.

However, the median  $\chi^2$ on the companion astrometry is slightly worse than for the solutions of the MCMC on the direct imaging alone, but not significantly. Moreover, \hipp \ and Gaia proper motion are well fitted except for the Gaia proper motion in RA that is not fitted because the HD 206893 B tangential velocity contribution in RA was null during Gaia observations.
We also find longer periods, $P$ between $35$ and $\SI{54}{years}$ with a maximum at $\SI{39}{years}$, which are only consistent at $2-\sigma$ with the periods found in Section \ref{POS}.
Adding \hipp \ and Gaia constraints favors lower eccentricities, with a maximum at $0.14$, and orbits misaligned with the star equator plane, about $i = 45^{+3}_{-3}$ from pole-on.
In order to fit \hipp \ and Gaia data, the MCMC selected results that fit the companion astrometry less well and therefore were discarded by the MCMC done in Section \ref{POS} and \ref{DI_RV}, which shows that \hipp \ and Gaia data do not probe the dynamical influence of the BD only.

\label{RV_ID_HG}

\section{Discussion}

\label{discussion}

Assuming that the RV drift in the RV of HD 206893 is due to HD 206893 B leads to a mass that is incompatible at $2 \sigma$ with the $12-\SI{50}{M_J}$ mass inferred from its observed photometry and spectra by \cite{Delorme} and with \hipp \ and Gaia proper motion measurements.

An additional inner body that would contribute significantly to the observed RV drift and the tangential velocity in RA is needed. Its orbital period should be larger than our time baseline of $\SI{1.6}{yr}$ ($\SI{1.4}{au}$). 
We used the following formula from \cite{Lazzoni} for the width of the chaotic zone around a planet to estimate the third body's semimajor axis $a_{crit}$ beyond which the system is not stable:
\begin{equation}
a_{crit}= a_B(1-e_B)(1-1.8(\mu e_B)^{\frac{1}{5}}),
\end{equation}
where $a_B$ is the first companion semimajor axis, $e_B$ is its eccentricity, and $\mu$ is the mass ratio of the first companion to the star. 
Using the characteristics of HD 206893 AB estimated by  \cite{Delorme}, $M_*=1.32 \ M_{\odot}$, $M_c=\SI{22}{M_J}$, $e=0.31$ and $a_B$ between $8$ and $\SI{10}{au}$, we obtain a critical semimajor axis between $2.1$ and $\SI{2.6}{au}$, which corresponds to periods between $3$ and $\SI{4}{yr}$ and to an angular separation on the sky between $50$ and $\SI{60}{\milli  as}$.
The critical semimajor axis can be underestimated because the presence of the inner companion induces a wobble of the host star, which produces a bias on the relative positions of the BD toward the host star and thus produces a bias on the eccentricity of its orbit \citep{Pearce}.
To constrain the properties of the inner body, we used a MCMC to fit the RV alone with a sinusoid ($e=0$) and for the inclination of the star: $i=\SI{30}{\degree}$ from pole-on, leaving free the phase, mass of the second companion, and period between $1.6$ and $\SI{4}{yr}$. The fit converges to a mass of $\SI{15}{M_J}$. Such a mass would lead to a contrast lower than $10^{-4}$ in the near infrared.
Given the expected contrast and the upper limit on the separation, it is not possible to detect this body with VLT \sphere.
This third body can be compatible with the  \hipp \ and Gaia constraint as its period is close to their time baseline; the impact of this third body on the proper motion of the star is then averaged.

\hipp \ and Gaia data imply a dynamical mass of $10^{+5}_{-4} \  \si{M_J}$ for HD 206893 B. This low mass is compatible with an age for the system lower than $\SI{50}{\mega yr}$. Its  probability of membership to the Argus association is higher than previously estimated. It would also explain the low luminosity of the BD compared to the models and its redness reported by \cite{Delorme}.
However, we note that this mass can be under or overestimated as the presence of the second companion could impact on the star tangential velocity.
More RV and direct imaging data are needed to further constrain both companions of HD 206893 system.

\begin{acknowledgements}
We acknowledge support from the French CNRS and from the Agence Nationale de la Recherche (ANR grant GIPSE ANR-14-CE33-0018). Part of the observations were performed under SHINE GTO time.
These results have made use of the SIMBAD database, operated at the CDS, Strasbourg, France. 
The authors thank Philippe Delorme and Eric Lagadec (SPHERE Data Center) for their help with the data reduction. We acknowledge financial support from the Programme National de Plan\'etologie (PNP) and the Programme National de Physique Stellaire (PNPS) of CNRS-INSU. This work has also been supported by a grant from the French Labex OSUG@2020 (Investissements d'avenir -- ANR10 LABX56). The project is supported by CNRS. This work has made use of the SPHERE Data Centre, jointly operated by OSUG/IPAG (Grenoble), PYTHEAS/LAM/CeSAM (Marseille), OCA/Lagrange (Nice), and Observatoire de Paris/LESIA (Paris). SPHERE is an instrument designed and built by a consortium consisting of IPAG (Grenoble, France), MPIA (Heidelberg, Germany), LAM (Marseille, France), LESIA (Paris, France), Laboratoire Lagrange (Nice, France), INAF -- Osservatorio di Padova (Italy), Observatoire astronomique de l'Universit\'e de Gen\`eve (Switzerland), ETH Zurich (Switzerland), NOVA (Netherlands), ONERA (France), and ASTRON (Netherlands), in collaboration with ESO. SPHERE was funded by ESO, with additional contributions from the CNRS (France), MPIA (Germany), INAF (Italy), FINES (Switzerland), and NOVA (Netherlands). SPHERE also received funding from the European Commission Sixth and Seventh Framework Programs as part of the Optical Infrared Coordination Network for Astronomy (OPTICON) under grant number RII3-Ct-2004-001566 for FP6 (2004--2008), grant number 226604 for FP7 (2009--2012), and grant number 312430 for FP7 (2013--2016).
This work has been supported by a grant from Labex OSUG@2020 (Investissements d’avenir – ANR10 LABX56).
C.\,P. acknowledges support from the ICM (Iniciativa Cient\'ifica Milenio) via the N\'ucleo Milenio de Formaci\'on Planetaria grant.
C.\,P. acknowledge financial support from Fondecyt (grant 3190691)
\end{acknowledgements}

\section{References}

\bibliography{206}

\bibliographystyle{aa}

\appendix

\onecolumn

\section{MCMC results on direct imaging astrometry}
 \label{result_pos}

\begin{figure*}[ht!]
\includegraphics[width=1\hsize, height =1\hsize ]{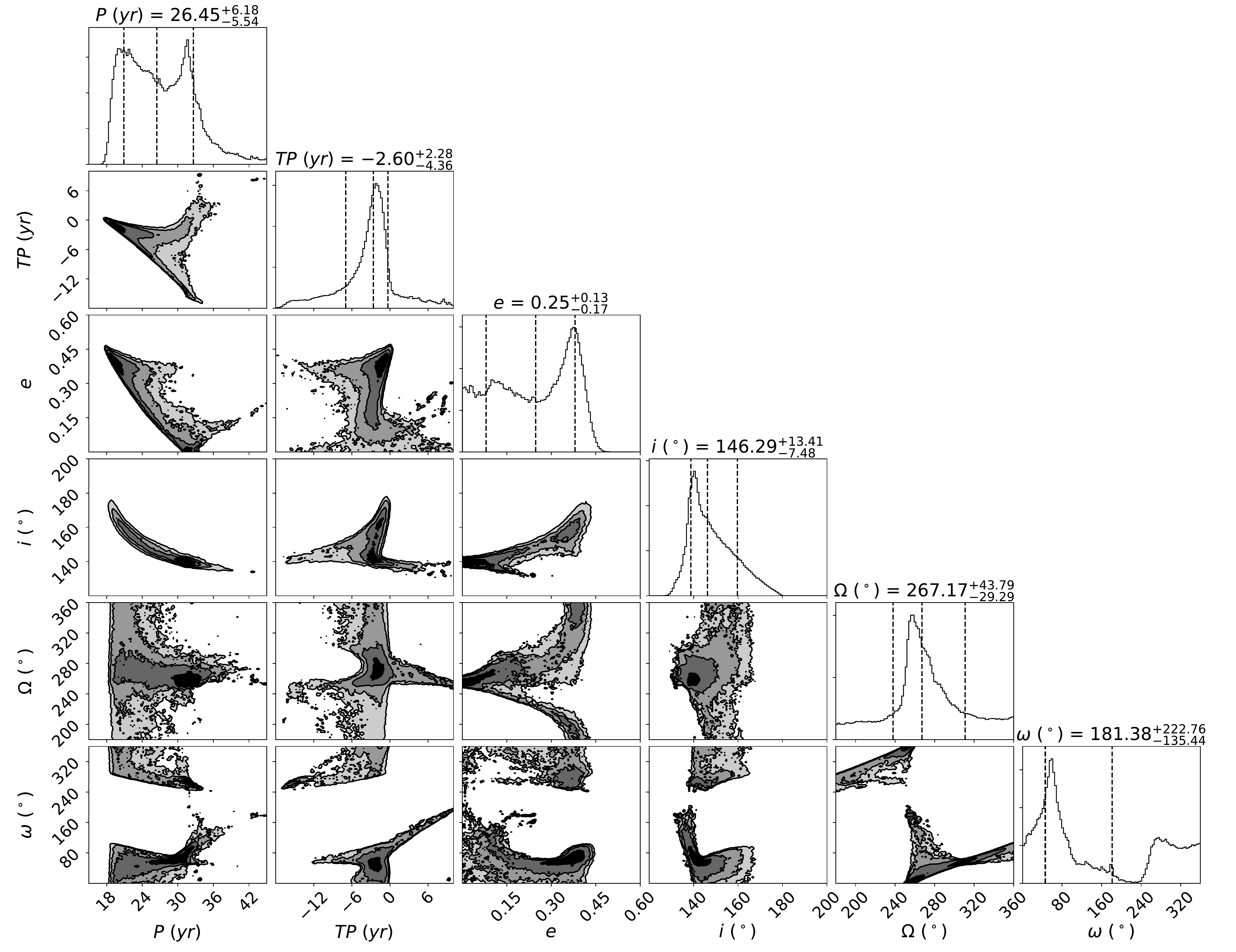}
\caption{Estimated probability distributions of the orbital parameters HD 206893 B obtained with the direct imaging astrometry data. \label{corner_pos}}
\end{figure*}

\twocolumn
\begin{figure}[ht!]
\centering
\includegraphics[width=1\hsize]{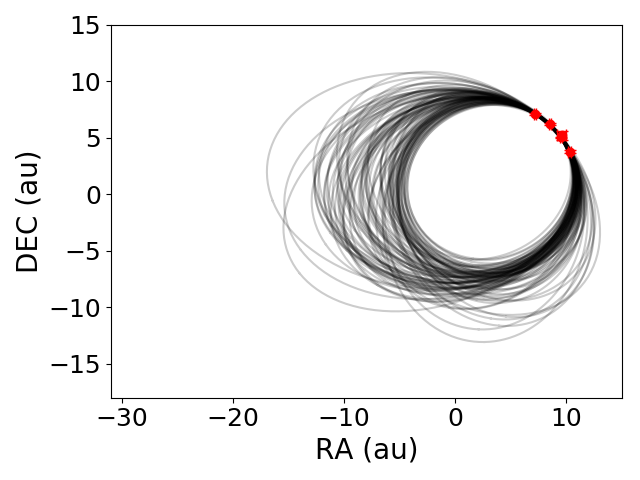}
\caption{HD 206893 B orbit solutions at the end of the MCMC on the direct imaging astrometry alone (\emph{black}) compared to the observations  (\emph{red}). \label{pos_pos}}
\end{figure}

\clearpage

\onecolumn
\section{MCMC results on direct imaging astrometry and RV}
 \label{result_pos_rv}

\begin{figure*}[ht!]
\includegraphics[width=1\hsize, height =1\hsize]{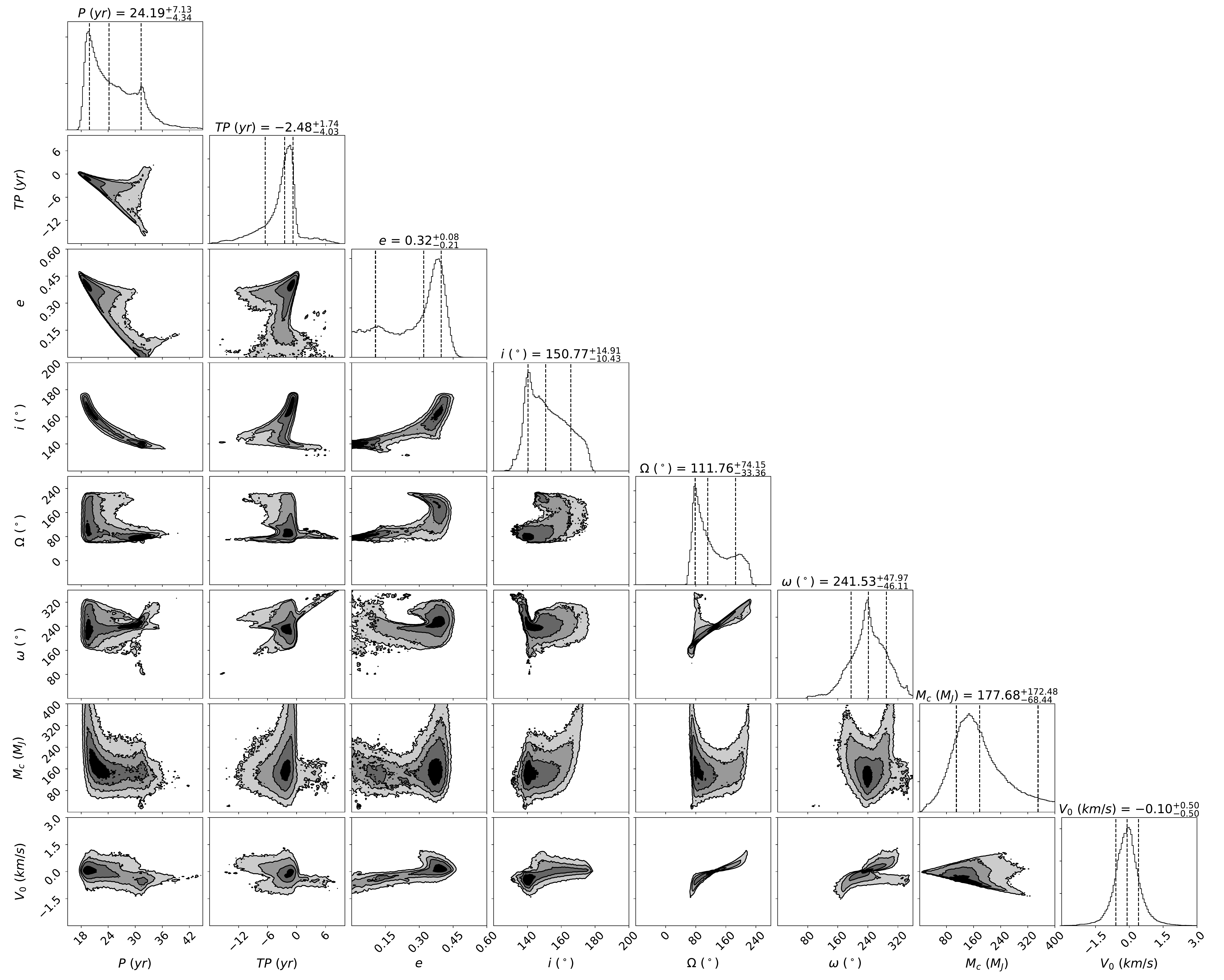}
\caption{Estimated probability distributions of the orbital parameters and RV parameters of HD 206893 B obtained with direct imaging astrometry data, and the RV drift estimated from the fit of the pulsations described in Section \ref{fit_puls}. \label{corner_corr_puls}}
\end{figure*}

\begin{figure}[ht!]
\centering
\includegraphics[width=0.45\hsize,valign=m]{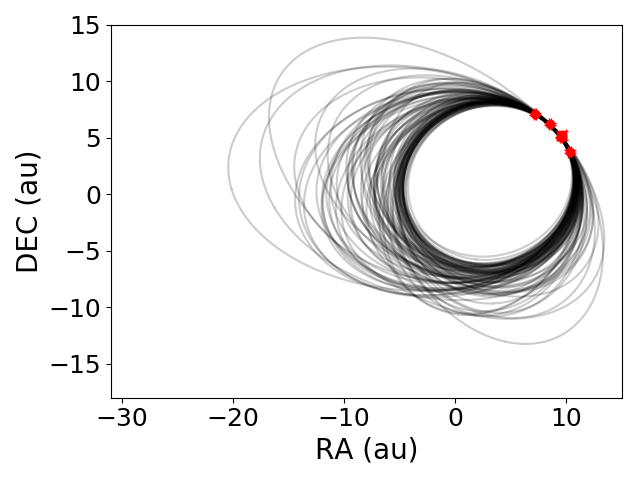}
\includegraphics[width=0.5\hsize,valign=m]{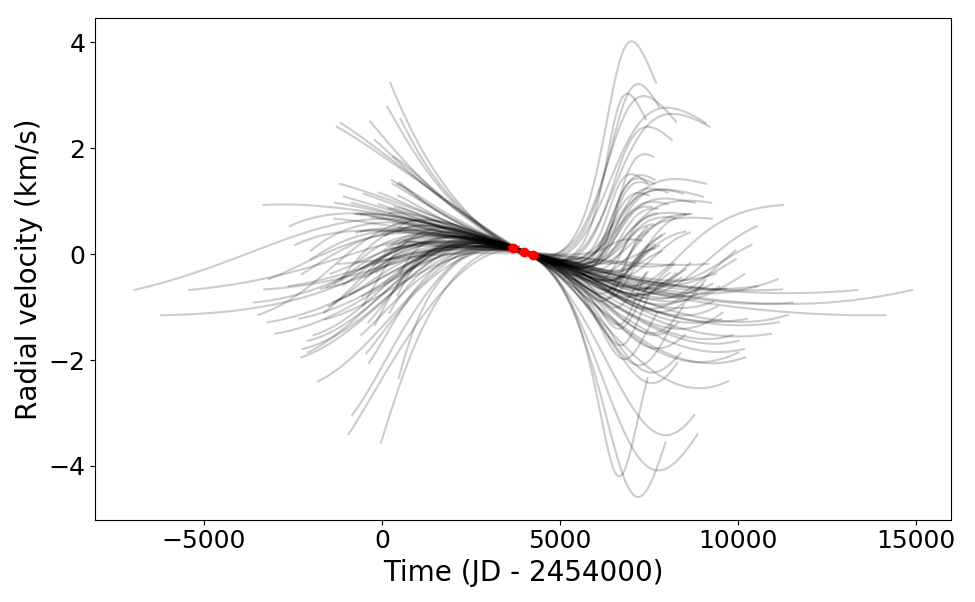}
\caption{HD 206893 B orbit and RV solutions at the end of the MCMC on direct imaging astrometry and RV (black)\ compared to the observations  (red). \label{pos_rv_pos_rv}}
\end{figure}

\begin{figure}[ht!]
\centering
\includegraphics[width=0.49\hsize,valign=m]{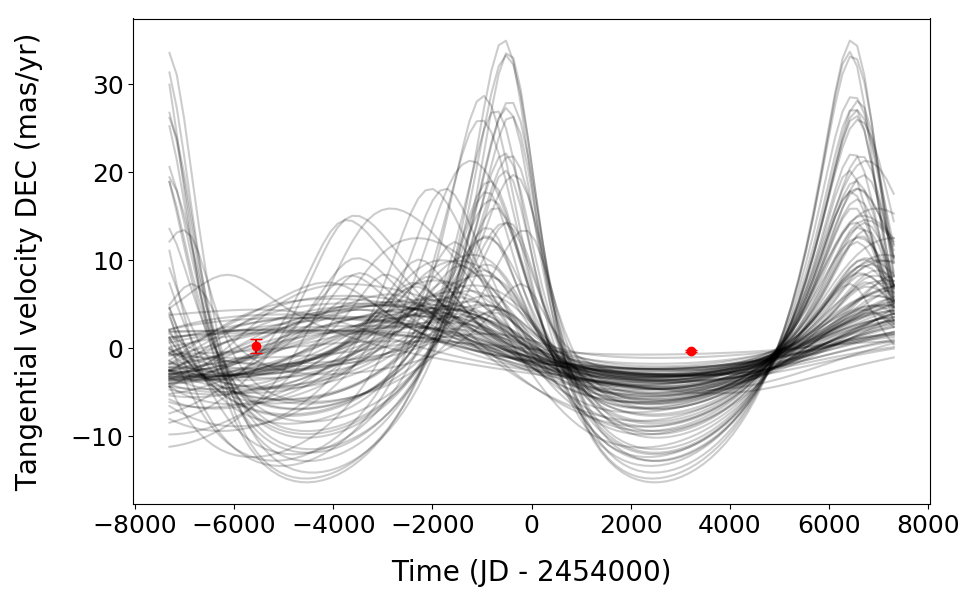}
\includegraphics[width=0.49\hsize,valign=m]{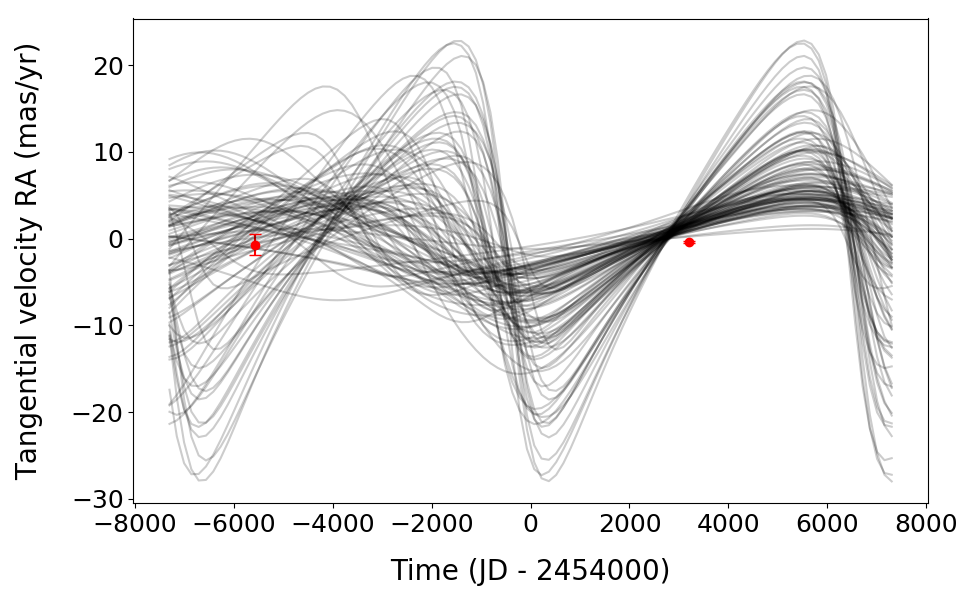}
\caption{HD 206893 B tangential velocity corresponding to the solutions at the end of the MCMC on direct imaging astrometry and RV (black) compared to the proper motion measured by \hipp \ and Gaia after removing the photocenter proper motion measured by \hipp \ and Gaia, presented at the mean time of their observations (red). \label{tg_pos_rv}}
\end{figure}

\clearpage
\section{MCMC results on direct imaging astrometry, RV, and proper motion}
\label{result_pos_rv_h_g_hg}

\begin{figure*}[ht!]
\includegraphics[width=1\hsize,height= 1\hsize]{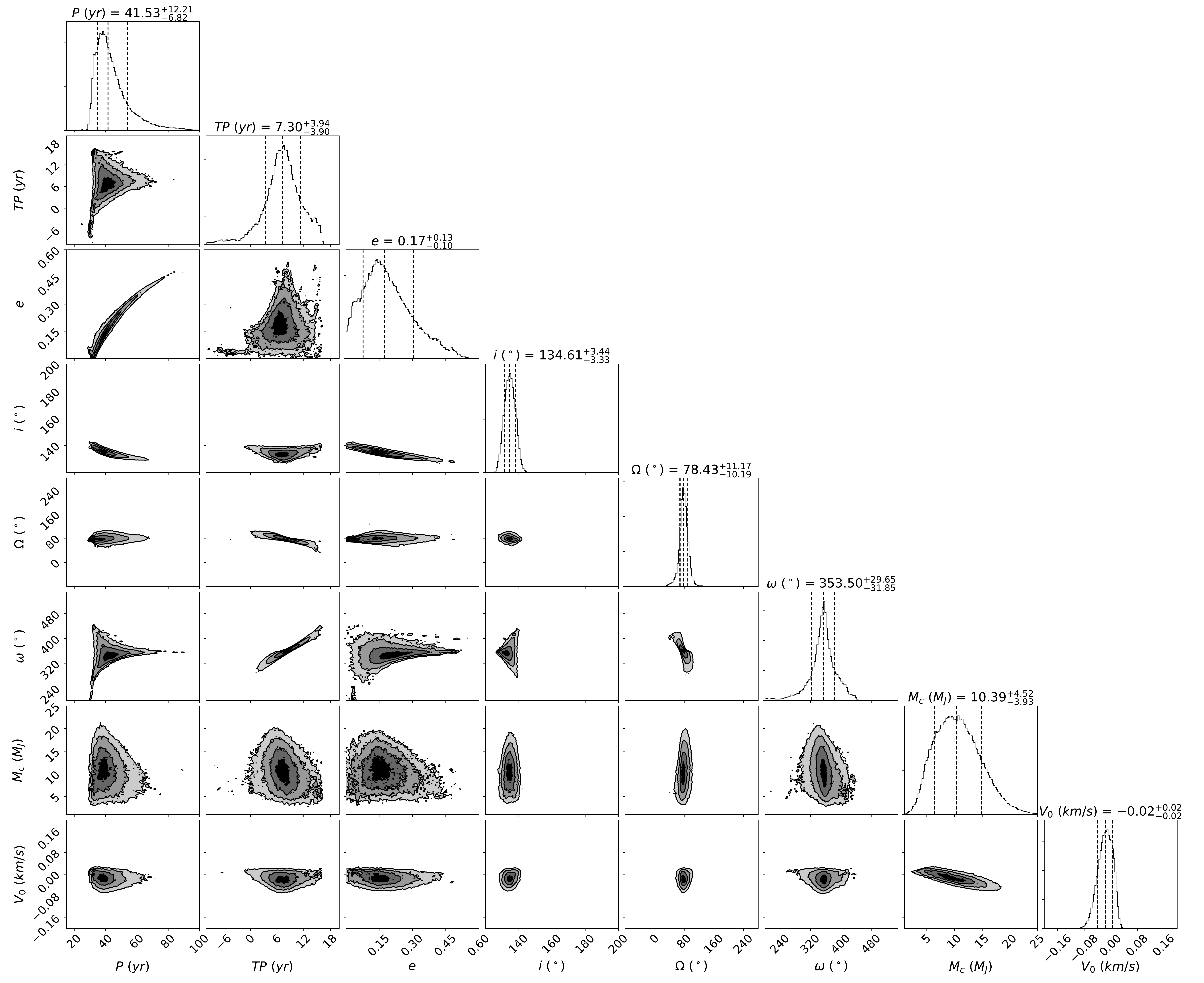}
\caption{Same as Figure \ref{corner_corr_puls}  adding proper motion constraints as described in Section \ref{g_h}.\label{corner_corr_hg}}
\end{figure*}

\begin{figure}[ht!]
\centering
\includegraphics[width=0.45\hsize,valign=m]{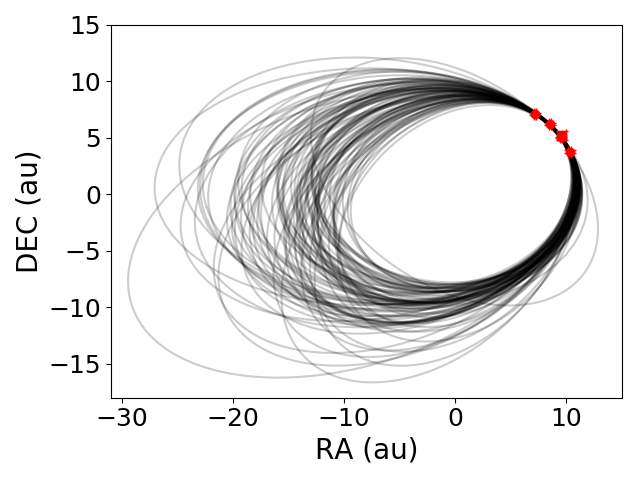}
\includegraphics[width=0.5\hsize,valign=m]{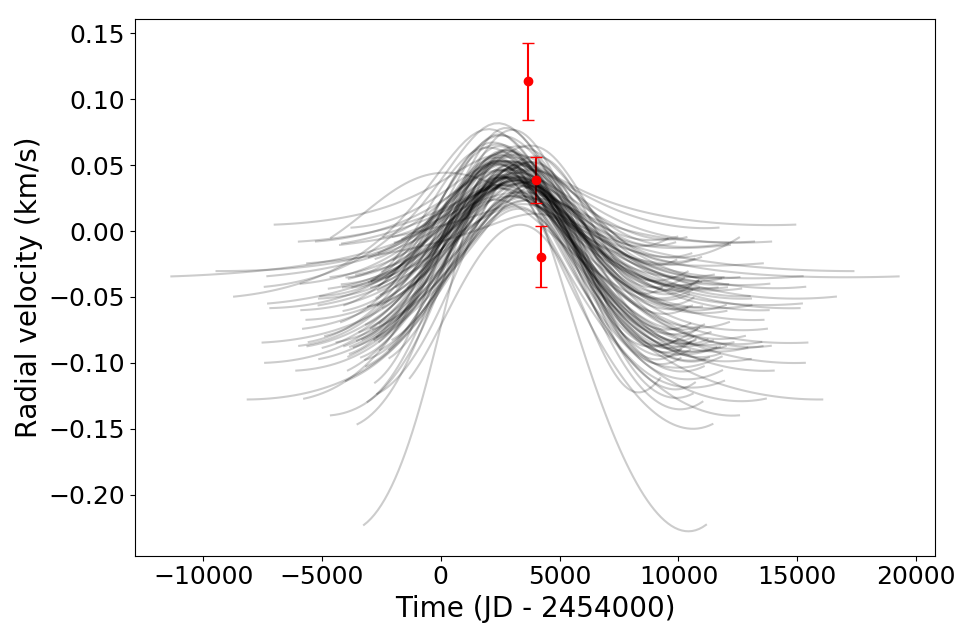}
\caption{HD 206893 B orbit and RV solutions at the end of the MCMC on direct imaging astrometry, RV, and proper motion (black) compared to the observations (red). \label{pos_rv_pos_rv_hg}}
\end{figure}

\begin{figure}[ht!]
\centering
\includegraphics[width=0.45\hsize,valign=m]{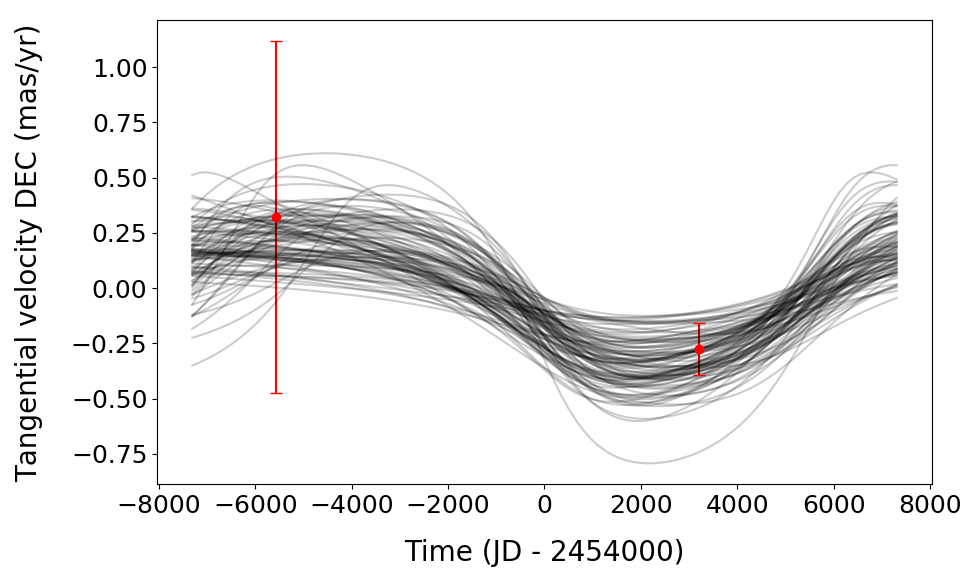}
\includegraphics[width=0.45\hsize,valign=m]{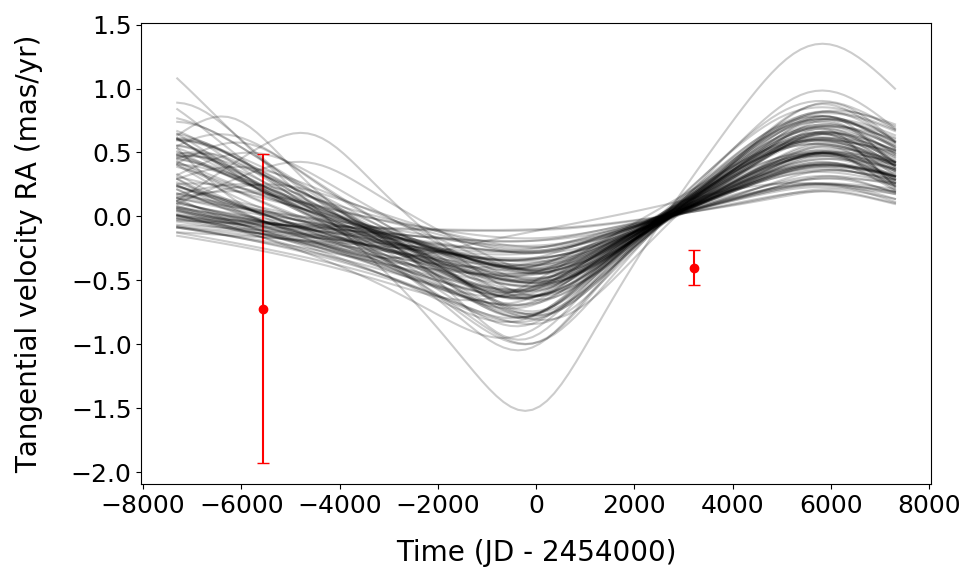}
\caption{HD 206893 B tangential velocity solutions at the end of the MCMC on direct imaging astrometry, RV, and proper motion (black) compared to the proper motion measured by \hipp \ and Gaia after removing the photocenter proper motion measured by \hipp \ and Gaia, presented at the mean time of their observations (red). \label{tg_pos_rv_hg}}
\end{figure}

\clearpage

\twocolumn

\section{\hipp \ and Gaia data exploitation}

\subsection{\hipp \ and Gaia data correction}

\label{corr_gaia}
We used \hipp-2 \citep{hipp2} and Gaia DR2 \citep{DR2A1} proper motion and position measurements of HD 206893 for our analysis.
First we verified that both \hipp \ and Gaia 5-parameter solutions are well behaved for HD 206893. 
The goodness of fit parameter $F2$ is $0.71$ for the \hipp \ reduction and the renormalized unit weight error\footnote{\url{https://www.cosmos.esa.int/web/gaia/dr2-known-issues}} (Hereafter $ruwe$)  is $1.1$ for Gaia, which ensures that the five-parameter solutions are relevant. 

Then we corrected the \hipp \ and Gaia proper motions measures from systematics. The global rotation between the Gaia and Hipparcos proper motions, which takes into account systematics on both sides \citep{Lindegren2018}, was corrected for $w_X=-0.011, w_Y=0.265, w_Z=0.127$. 
We quadratically added an extra error of $\SI{0.5}{\milli as \per yr}$ on \hipp \ proper motion errors and we rescaled the Gaia DR2 proper motion errors and the errors on the proper motion deduced from the difference of position by a factor $1.4$.
The rotations and errors inflations have been derived from the distributions of well behaved Gaia DR2 and \hipp \ sources (\emph{e.g.,} sources not already flagged as binaries in Simbad, with an Hipparcos $|F2|<5$, a Gaia $ruwe<1.4$ and proper motions compatible within $5\sigma$). However, unresolved binaries are still expected to be present with this selection and our error inflations are therefore conservative. 

\subsection{Barycenter proper motion determination}

\label{ref_pm}

To compute the proper motion of the barycenter of the system we must correct the proper motion deduced from the difference of position between \hipp \ and gaia measurements (photocenter proper motion) by the effect of the star position induced by the companion.

The position of the star induced by the companion  compared to the barycenter of the system on the sky at a time $t$ is 

\begin{align}
\nonumber
&X_*(t) =  -X_c (t)*\frac{M_c}{(M_c+M_*)d}\\ \nonumber
&Y_*(t) = -Y_c (t)*\frac{M_c}{(M_c+M_*)d},\\ \nonumber
\end{align}

where $X_c(t)$ and $Y_c (t)$ are the coordinates of the companion on the sky in $DEC$ and $RA$, respectively, $M_c$ and $M_*$ are the mass of the companion and mass of the star; $d$ is the distance of the star in $\si{pc}$.

As the measurements of  \hipp \ and Gaia DR2 are each based on a two-year timespan, we computed the mean of the positions induced by the companion compared to the barycenter of the system over the corresponding timespans.

We used the times of \hipp \ observations available in the \hipp-2 catalog \citep{hipp2}.
The specific times of observations of Gaia are not published yet. We then took $62$ times that were equally spaced between the start and the end of Gaia measurements, i.e., 25 July 2014 and 23 May 2016, respectively  \citep{DR2A1}.

The barycenter proper motion is then
\begin{align}
\nonumber
&\mu_{Ref}^{RA} = \mu_{HG}^{RA}-\frac{ <Y_*>_{G}- <Y_*>_{H}}{<t_{Hipp}>-<t_{Gaia}>}\\ \nonumber
&\mu_{Ref}^{DEC} = \mu_{HG}^{DEC}-\frac{ <X_*>_{G}- <X_*>_{H}}{<t_{Hipp}>-<t_{Gaia}>}.\\ \nonumber
\end{align}

\subsection{Proper motion induced by the companion model}

\label{hg_model}

We present our model to compute the proper motion induced by the companion at \hipp \ and Gaia measurement.
First, we computed the orbital speed of the companion in the plane of its orbit at a time $t$ (in $\si{au \per yr}$) as follows:

\begin{displaymath}
\begin{array}{lc}
V_c^X(t) = -\frac{2 \pi a^2 \sin{E}}{Pr(t)}\\
V_c^Y(t) =  \frac{2 \pi a^2 \cos{E}}{Pr(t)}\sqrt{(1-e^2)},
\end{array}
\end{displaymath}

where $P$ is the period in year, $a$ the semimajor axis in $\si{au}$, the eccentricity $e,$ and the eccentric anomaly $E$ of the orbit.

The orbital speed is then projected on the plane of the sky as follows:
\begin{align}
\nonumber
&V_c^{RA}(t) & =&\qquad V_c^X(t)*(\cos{\omega}\sin{\Omega}+\sin{\omega}\cos{i}\cos{\Omega}) \\ \nonumber
&&&+V_c^Y(t)*(-\sin{\omega}\sin{\Omega}+\cos{\omega}\cos{i}\cos{\Omega}) \\\nonumber
&V_c^{DEC}(t) &  =&\qquad V_c^X(t)*(\cos{\omega}\cos{\Omega}-\sin{\omega}\cos{i}\sin{\Omega}) \\ \nonumber
&&&+V_c^Y(t)*(-\sin{\omega}\cos{\Omega}-\cos{\omega}\cos{i}\sin{\Omega}), \\\nonumber
\end{align}

where $\omega$  is the argument of periapsis in radian,  $\Omega$ the longitude of the ascending node in radian, and $i$ the inclination of the orbit in radian.

Finally, the star proper motion induced by the companion in $\si{as \per  yr}$ is

\begin{align}
\nonumber
&\mu_*^{RA}(t)  = -V_c^{RA}(t)*\frac{M_c}{(M_c+M_*)d} \\ \nonumber
&\mu_*^{DEC}(t)  = -V_c^{DEC}(t)*\frac{M_c}{(M_c+M_*)d}, \\ \nonumber
\end{align}

where $M_c$ and $M_*$ are the mass of the companion and the mass of the star and $d$ is the distance of the star in $\si{pc}$.

As the measurements of  \hipp \ and Gaia DR2 are each based on a two-year timespan we computed the mean of the theoretical proper motions over the corresponding timespans.

We used the times of \hipp \ observations available in the \hipp-2 catalog \citep{hipp2}.
The specific times of observations of Gaia are not published yet. We then
took  $62$ times that were equally spaced between the start
and the end of Gaia measurements, i.e., 25 July 2014 and 23 May 2016, respectively  \citep{DR2A1}.

\section{MCMC inputs}
\label{MCMC description}

\subsection{MCMC on the direct imaging astrometry}

\label{desc_pos}

We modeled the observations by a Keplerian orbit projected on the sky.
The likelihood function is a centered and reduced Gaussian of the $\chi^2$ function between the model and the observations. The logarithm of our likelihood function is

\begin{displaymath}
ln  \mathcal{L} = -\frac{1}{2} \left( \sum_{i=1}^{N_{DI}}\left( \frac{y^{DI}_i-f_i^{DI}(X)}{\sigma_i^{DI}}\right)^2\right),
\end{displaymath}

where $y$ are the observations, $\sigma$ their corresponding variances, and$f(X)$ the model for parameters X.

\bigbreak

To facilitate the convergence of the MCMC we chose the following parameters: 

\begin{itemize}
\item $logP$: the decimal logarithm of the period in years;
\item  $f_{TP}$: a factor for which the time at the periastron is computed, i.e., $TP = f_{TP}*P$. In this study the $TP$ is in years and corresponds to the closest periastron passage from the first direct imaging epoch;
\item $e$: the eccentricity;
\item $i$: the inclination in radian, $\SI{0}{\degree}$ and $\SI{90}{\degree}$ correspond to pole-on and edge-on orbit, respectively;
\item $\Omega$: the longitude of the ascending node in radian;
\item  $\omega$: the argument of periapsis in radian.
\end{itemize}
We used the same model as \cite{Model_POS_RV}, which works on a combination of the parameters presented above.

\bigbreak
We used uniform priors in the following intervals:
\begin{itemize}
\item $0.5 <  logP \leq 3$;
\item $ -0.5 \leq f_{TP} \leq 0.5$;
\item $0 < e < 1$;
\item $ -2 \pi < \Omega < 2 \pi$;
\item $ 0 < \omega < 4 \pi$.
\end{itemize}

Elsewhere, the prior is null. For the inclination, we took a prior that is the product of a uniform prior in $cos(i)$ and a uniform prior in inclination in the range  $0 < i < \pi$.

\subsection{MCMC on the direct imaging astrometry and RV}

\label{desc_pos_rv}

We added to the model a Keplerian in RV to model both the astrometry of the companion and the RV.
The likelihood function becomes\begin{align}
\nonumber
ln  \mathcal{L} = &-\frac{1}{2} \left( \sum_{i=1}^{N_{RV}}\right.\left( \frac{y^{RV}_i-f_i^{RV}(X)}{\sigma_i^{RV}}\right)^2 \\ \nonumber
 &+ \left. \sum_{i=1}^{N_{DI}}\left(\frac{y^{DI}_i-f_i^{DI}(X)}{\sigma_i^{DI}}\right)^2\right),
\end{align}

where $y$ are the observations, $\sigma$ their corresponding variances, and$f(X)$ the model for parameters X.

\bigbreak

To facilitate the convergence of the MCMC we chose additional parameters as

\begin{itemize}
\item $M_c$: the mass of the companion in $\si{M_J}$;
\item $f_{V_0}$: a factor for which the offset in RV are computed, $V_0=f_{V_0}*K(M_c,P,e)$, with $K$ the semi-amplitude of the RV.
\end{itemize}
We used the same model as \cite{Model_POS_RV}, which works on a combination of the parameters presented above.

\bigbreak
We used uniform priors for the additional parameters in the following intervals:
\begin{itemize}
\item $1 < M_c \leq 1000$
\item $ -1 \leq f_{V_0} \leq 1$.
\end{itemize}

Elsewhere, the prior is null.

\subsection{MCMC on the direct imaging astrometry and RV and proper motion}

\label{desc_pos_rv_hg}

We added to the model the tangential velocity model described in \ref{hg_model} and the barycenter proper motion model described in \ref{ref_pm} to fit the difference between the proper motion measured by \hipp \ and Gaia and the  barycenter proper motion. 
The likelihood becomes 

\begin{align}
\nonumber
ln  \mathcal{L} = -\frac{1}{2} &\Biggl(\sum_{i=1}^{N_{RV}}\left( \frac{y^{RV}_i-f_i^{RV}(X)}{\sigma_i^{RV}}\right)^2\\ \nonumber
&+ \sum_{i=1}^{N_{DI}}\left( \frac{y^{DI}_i-f_i^{DI}(X)}{\sigma_i^{DI}}\right)^2  \\ \nonumber
 &+  \biggl(\frac{(y^{HG}-y_{ref}^{HG})-f^{HG}(X)}{ \sqrt{ \sigma^2_{HG} +\sigma^2_{HG,ref} } }\biggr)^2\Biggr),
\end{align}

where $y$ are the observations, $\sigma$ their corresponding variances, and $f(X)$ the model for parameters X.

\bigbreak

We used the parameters and prior described in Appendix \ref{desc_pos_rv}.

\subsection{Description of the MCMC runs}

We used \emph{emcee} with the prior and likelihood function as described in Appendix \ref{desc_pos} and  Appendix \ref{desc_pos_rv}.

In order to facilitate the convergence, we ran two MCMC for each analysis described in Sections \ref{RV_ID}, \ref{corr} and \ref{RV_ID_HG}.
We ran the first MCMC with 100 walkers that were initialized  randomly with a uniform probability distribution for each parameter in the intervals of their prior for $100\,000$ iterations.
Then a second MCMC was performed with 100 walkers initialized with a uniform probability distribution for each parameter in the intervals where the majority of the walkers of the first run were converging.
This approach permits us to check the uniqueness of the solution.

To ensure convergence, as our problem show a multimodal posterior, we verified that the number of iterations was greater than ten times the autocorrelation time of the chain for each parameters \citep{emcee}.

\end{document}